\documentclass[reprint,
 amsmath,amssymb,
 aps,
pra,
prl,
rmp,
prstab,
prstper,
floatfix,
]{revtex4-1}

\usepackage{graphicx}
\usepackage{filecontents}
\usepackage{tabularx}
\usepackage{appendix}
\usepackage{dcolumn}
\usepackage{bm}
\DeclareRobustCommand{\rchi}{{\mathpalette\irchi\relax}}
\newcommand{\irchi}[2]{\raisebox{\depth}{$#1\chi$}}

\begin{document}

\preprint{APS/123-QED}

\title{Bi kagome sublattice distortions by quenching and flux pinning in superconducting RbBi$_2$}

\author{Sharon S. Philip}
\affiliation{University of Virginia, Charlottesville, VA 22904, USA}
\author{Junjie Yang}
\affiliation{University of Virginia, Charlottesville, VA 22904, USA}
\author{P. F. S. Rosa}
\affiliation{Los Alamos National Laboratory, Los Alamos, NM 87545, USA}
\author{J. D. Thompson}
\affiliation{Los Alamos National Laboratory, Los Alamos, NM 87545, USA}
\author{K. L. Page}
\affiliation{The University of Tennessee, TN , USA}
\author{Despina Louca*}
\affiliation{University of Virginia, Charlottesville, VA 22904, USA}

\date{\today}

\begin{abstract}
 The properties of RbBi\textsubscript{2}, a 4.15 K superconductor, were investigated using magnetic field, pressure and neutron diffraction. Under hydrostatic pressure, an almost 50 \%\ reduction of T\textsubscript{c} is observed, linked to a low Debye temperature estimated at 165 K. The resistivity and magnetic susceptibility were measured on quenched and slow-cooled polycrystalline samples. The resistivity follows a low temperature power-law dependence in both types of samples, while the diamagnetic susceptibility, $\chi$, is dependent on the sample cooling history. Slow-cooled samples have a $\chi= -1$ while quenched samples have $\chi< -1$ due to grain size differences. Evidence of the effects of the cooling rate is also discerned from the local structure, obtained by neutron diffraction and the pair density function analysis. Slow-cooled samples have structurally symmetric Bi hexagons, in contrast to quenched samples in which disorder is manifested in periodic distortions of the Bi hexagonal rings of the kagome sublattice. Disorder may lead to flux pinning that reduces vortex mobility, but T\textsubscript{c} remains unaffected by the cooling rate.  
\end{abstract}

\maketitle

\section{\label{sec:intro}Introduction}
Superconductivity, a state characterized by zero electrical resistance, has been observed in many classes of materials, from simple metals, amorphous and granular systems to complex ladder structures~\cite{london1954superfluids, bardeen1957theory, waldram1996superconductivity, tinkham2004introduction, de2018superconductivity, ginzburg2009theory, collver1973superconductivity, deutscher1974superconducting, takahashi2008superconductivity, dagotto1992superconductivity, dagotto1996surprises, dagotto1999experiments}, and has been the subject of intense debate for over a century. Bismuth (Bi) based superconductors attracted a lot of attention in the last decade or so in part because of the possible interplay of topology and superconductivity. Bi exhibits strong spin-orbit (SO) coupling linked to topologically nontrivial band structures in its compounds. In topological materials, the quantum state is entangled to an extent where its emergent quasiparticles exhibit exotic behaviors that are unique, and cannot be reproduced in conventional solids. These exotic properties are topologically protected as they are robust against symmetry-preserving perturbations. Several binary Bi compounds are currently under investigation for topological superconductivity. Some of the known bismuth-alkali and alkaline-earth metal intermetallic compounds studied for superconductivity are LiBi, NaBi, KBi\textsubscript{2}~\cite{reynolds1950superconducting, sun2016type}, CsBi\textsubscript{2}~\cite{roberts1976survey}, Ca\textsubscript{11}Bi\textsubscript{10-x}~\cite{sturza2014superconductivity}, CaBi\textsubscript{2}~\cite{merlo1994crystal,dong2015rich}, CaBi\textsubscript{3},  SrBi\textsubscript{3}~\cite{matthias1952search}, BaBi\textsubscript{3}~\cite{matthias1952search} and Ba\textsubscript{2}Bi\textsubscript{3}~\cite{iyo2014superconductivity}. Among these, KBi\textsubscript{2} and CaBi\textsubscript{2} are reported to show type-I superconductivity. While the topological nature of their superconductivity has not yet been confirmed, it is important to explore their materials properties. In this work, we report on the transport, magnetization and structural properties of RbBi\textsubscript{2} alkali bismuth superconductor.

Although discovered decades ago, little is known of the properties of  RbBi\textsubscript{2}, which is isostructural to KBi\textsubscript{2}~\cite{sun2016type} with a superconducting transition temperature, T\textsubscript{c}, of 4.15 K. The resistivity is linear at high temperatures but follows a power law temperature dependence~\cite{chan2014plane,mckenzie1998violation} upon cooling below 25 K. Under pressure, strong suppression of T$_c$ is observed that is not typical of metals. By varying the sample cooling rate, we examined the diamagnetic susceptibility response which shows perfect diamagnetism in samples with slow cooling rates. Quenching leads to a $\chi < -1$. Larger grain size is expected with slow cooling rates, that increases the magnetic susceptibility because grain boundaries are reduced. The effects of quenching can also be seen in the atomic structure. Local distortions are observed in the Bi kagome rings of fast-cooled samples where the Bi-Bi bond lengths are no longer equivalent. By comparison, kagome rings of slow cooled samples are symmetric. Other systems can be found in the literature where varying the cooling rate can affect their superconducting and structural properties. For instance, the effects of annealing were previously studied in the FeTe\textsubscript{1-x}Se\textsubscript{x}~\cite{louca2011suppression} superconductor. In this system, the  suppression of T\textsubscript{c} was linked to changes in the chalcogen ion's z - parameter although the diamagnetic response remained the same. In the current system, although T\textsubscript{c} does not change, the diamagnetic response does.

\section{\label{sec:mat&meth}Materials and Methods}

The RbBi\textsubscript{2} ingots were prepared by solid state reaction, by mixing Rubidium pieces (Alfa Aesar, 99.5\%) and Bismuth powder (Alfa Aesar, 99.99\%) in a 1:2 molar ratio in an Argon filled glovebox. The quartz ampule with the combined mixture was vacuum sealed without exposure to air. The samples were heated at 700\textdegree C for 24 hours. Two batches of samples were synthesized, one that was slowly cooled down (furnace-cooled) to room temperature and the other quenched from 700\textdegree C in liquid N\textsubscript{2} to room temperature. The samples are denoted as AG (as-grown) and Q (quenched), respectively. The air sensitive samples were stored in a glovebox to avoid decomposition. The PPMS resistivity puck and samples for susceptibility measurements were prepared inside the glovebox and carefully transferred to the sealed chamber without exposure to air. Samples for the neutron diffraction measurements were loaded in the Vanadium can inside a glovebox sealed with He exchange gas. Electrical transport and magnetization measurements were performed as a function of magnetic field. High pressure resistivity measurements were also carried out at pressures ranging from 0 to 12 kbar. The time-of-flight (TOF) neutron diffraction experiment was carried out at the Nanoscale Ordered Materials Diffractometer (NOMAD) at the Spallation Neutron Source (SNS) of Oak Ridge National Laboratory (ORNL). The data were analyzed using the Rietveld refinement that yields the basic structural parameters for the periodic structure ~\cite{toby2001expgui} and the pair distribution function (PDF) analysis technique that provides information on the local arrangement of atoms without the assumption of periodicity. The PDF analysis was performed on the same neutron diffraction data used for the Rietveld refinement. NOMAD is a diffractometer with a large bandwidth of momentum transfers, $Q$, and provides the total structure function $S(Q)$. The $S(Q)$ was Fourier transformed into real-space as shown in Eqn.~\ref{eq:Grexp} to obtain the $G(r)$~\cite{warren1990x, pdfgetnpeterson2000}. The instrument background and empty sample can were subtracted from the $S(Q)$ and the data were normalized by vanadium. A maximum $Q$ of 40 Å\textsuperscript{-1} was used. The $G(r)$ corresponds to the probability of finding a particular pair of atoms with an inter-atomic distance $r$~\cite{egami2003underneath}.
\begin{eqnarray}
G(r)_{exp} = \frac{2}{\pi} \int_{0}^{\infty} Q[S(Q) - 1]sin(Qr)dQ
\label{eq:Grexp}.
\end{eqnarray}
The Debye temperature, $\theta_D$, was extracted from fitting the width of the PDF peaks assuming a correlated Debye model~\cite{athauda2015crystal,beni1976temperature,jeong2003lattice}. Using eqn.~\ref{eq:ThetaD}, the full width half maximum (FWHM), $\sigma_{ij}$, is extracted from the first PDF peak in RbBi\textsubscript{2} corresponding to the Bi-Bi nearest neighbour correlation. 
Here the Debye wavevector is given by $k_D = (6\pi^2N/V)^{1/3}$ where $N/V$ is the number density of the crystal; the Debye cutoff frequency $\omega_D = k_B\theta_D$, where $k_B$ is the Boltzmann constant; and $\Phi_n = \int_{0}^{\theta_D/T} x^n(e^x - 1)^{-1} dx$ where x is a dimensionless integration variable.

\begin{multline}
\sigma_{ij}^2 = \frac{6\hbar}{M\omega_D}[ \frac{1}{4} + \left(\frac{T}{\theta_D}\right)^2 \Phi_1 - \frac{1 - cos(k_{D} r_{ij})}{2(k_{D}r_{ij})^2} \\   - \left(\frac{T}{\theta_D}\right)^2 \int_{0}^{\frac{\theta_D}{T}} \frac{sin\left(\frac{k_{D} r_{ij}Tx}{\theta_D}\right)/ \left(\frac{k_{D} r_{ij}T}{\theta_D}\right)}{e^x - 1}dx ]
\label{eq:ThetaD}
\end{multline}

\begin{figure}[h!]
 \includegraphics[width = 85mm]{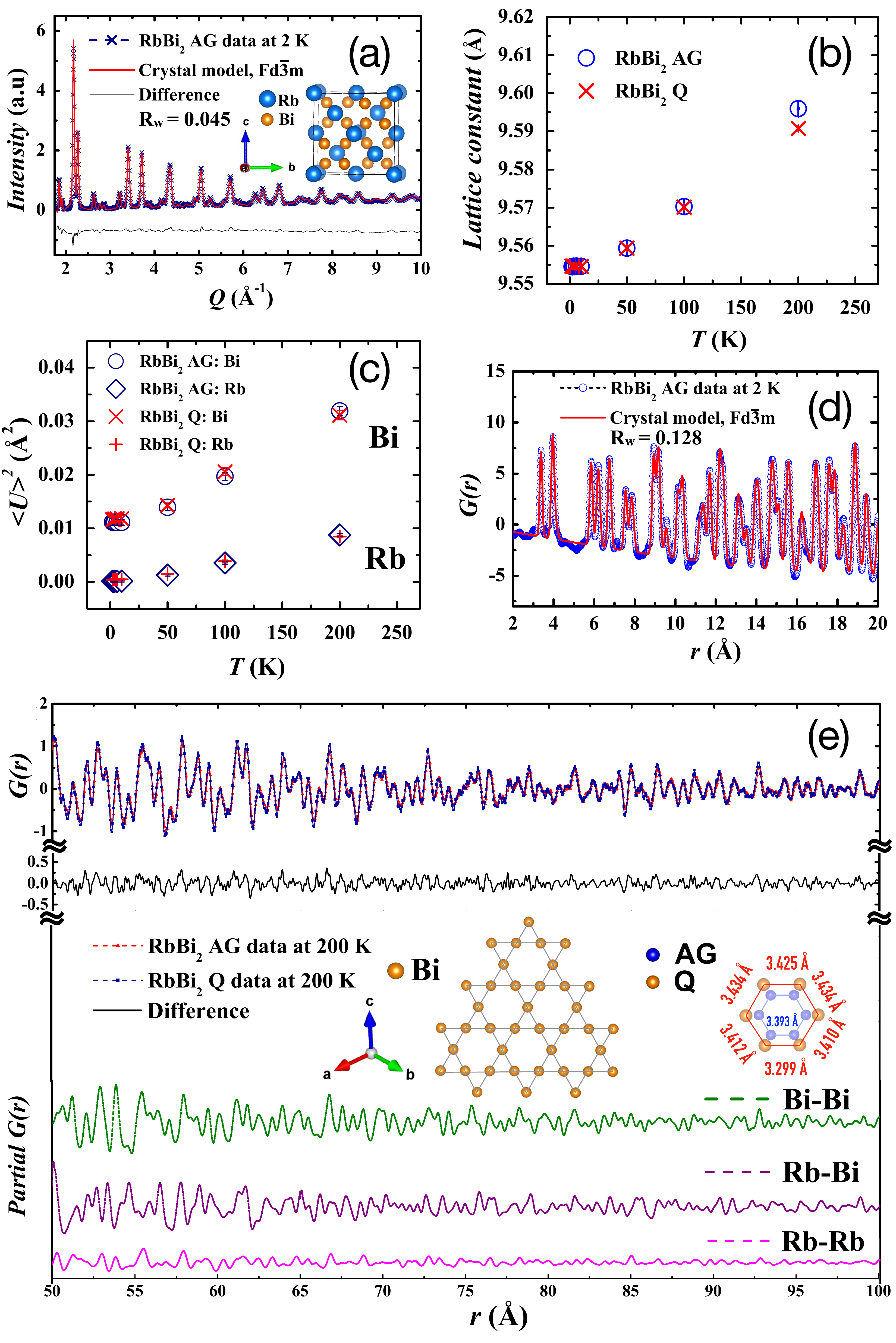}
\caption[width = \textwidth]{(a) The neutron powder diffraction pattern of as-grown (AG), slow-cooled RbBi$_2$ collected at 2 K is fit with a cubic
model that fits the data quite well. Shown in the inset is a model of the crystal structure. (b) The temperature dependence of the lattice constant determined from the Rietveld refinement. (c)  The isotropic thermal factors for Rb and Bi in the AG RbBi\textsubscript{2}. (d) The G(r) function determined for AG RbBi\textsubscript{2} at 2 K is compared to a model G(r) with the $Fd\overline{3}m$ symmetry. (e) A comparison of the G(r) for AG and quenched (Q) samples of RbBi\textsubscript{2} obtained from the diffraction data collected at 200 K and plotted from 50 to 100 {\AA}. The difference curve is shown below the data. The differences, although small, are attributed to the Bi sublattice distortions shown in the middle of the figure. This is deduced from the partial PDFs of the three pair correlations, Bi-Bi, Bi-Rb and Rb-Rb that are plotted below. These partials are determined from the model fitting of the 200 K AG data.}
\label{fig:characterization_structure}
\end{figure}

\section{\label{sec:results}Results and Discussion}
\subsection{\label{sec:results_a}Structural analysis}

ABi\textsubscript{2} is isostructural to the cubic MgCu\textsubscript{2} Laves phase, with $Fd\overline{3}m$ space group symmetry. The crystal structure is shown in the inset of Fig.~\ref{fig:characterization_structure}(a) where each unit cell consists of 8 Rb atoms and 16 Bi atoms. Four Bi atoms form a tetrahedron and the tetrahedra connect with each other by vertex-sharing to form a three-dimensional network. Meanwhile, the Rb atoms are arranged in a diamond sublattice which is intertwined with the network of Bi tetrahedra. The Bi sublattice forms a hyperkagome structure shown in the inset of Fig.~\ref{fig:characterization_structure}(e). In Fig.~\ref{fig:characterization_structure}(a) the $S(Q)$ for the AG sample obtained from the diffraction data collected at 2 K is plotted as a function of Q. The $S(Q)$ is compared to a model calculated based on the $Fd\overline{3}m$ symmetry. The fitting yields an Rw = 0.045. Similarly, the diffraction pattern collected for the Q-sample yields an equally good fitting with the same symmetry. Within the resolution of the NOMAD instrument, almost no differences can be discerned in the Bragg structure between the AG and Q samples.  In Figs.~\ref{fig:characterization_structure}(b) and~\ref{fig:characterization_structure}(c), the lattice constants ($a$ = 9.586{\AA}) and thermal factors obtained from the Rietveld refinement are plotted as a function of temperature for both the AG and Q samples. The lattice constant increases with increasing temperature, as expected. Note that while both samples show the same temperature dependence in their lattice constants and thermal factors, the $<U>^2$ displacement for Bi is significantly larger than the one for Rb, signifying larger thermal fluctuations at the Bi sites. 

Shown in Fig.~\ref{fig:characterization_structure}(d) is the $G(r)$ corresponding to the local atomic structure for the AG sample at 2 K. It is obtained by Fourier transforming the $S(Q)$ of Fig.~\ref{fig:characterization_structure}(a). Also shown in this figure is a comparison with a model $G(r)$ calculated from the atomic coordinates and unit cell dimensions of the periodic cubic cell  with the $Fd\overline{3}m$ symmetry. It can clearly be seen that up to 20 {\AA}, the cubic model fits the local atomic structure quite well and the agreement between the model and experiment is excellent. In this range, no differences can be detected between the atomic structures of the AG and Q samples either. However, when extending the $G(r)$ beyond 20 {\AA}, small differences become discernible between Q and AG, indicating that the structural changes manifested with quenching are becoming apparent above 20 {\AA}. Furthermore, the differences become more apparent at elevated temperatures. 

Fig.~\ref{fig:characterization_structure}(e) is a plot of the $G(r)$'s corresponding to the local atomic structures for Q and AG samples at 200 K in the 50 - 100 {\AA} range. The difference curve is shown below the data. Although the two samples are quite similar on average, small differences are detected in the  $G(r)$'s up to 100 {\AA}. A similar difference curve is observed from 20 to 50 {\AA}. What is the origin of these differences between the two samples? Shown in Fig.~\ref{fig:characterization_structure}(e) are the calculated partial functions obtained from the model used to fit the AG data at 200 K. The most significant contribution in the $G(r)$ arises from Bi correlations. While the Rb-Rb correlations are significantly reduced above 50 {\AA}, the Bi correlations are much stronger and contribute the most to the G(r) function. Thus the differences in the local structure observed between the AG and Q samples are primarily from the Bi kagome lattice. Shown in the inset of Fig.~\ref{fig:characterization_structure}(e) is a comparison of the Bi hexagon obtained from the real-space refinement of the AG and Q data at 200 K. We find that in the AG sample, the hexagon is symmetric with equal bond lengths (blue inner hexagon), while in the Q sample, the Bi hexagon is not symmetric and all bond lengths are slightly different (red outer hexagon). This is shown in the figure although not drawn to scale. Thus the quenching process results in a periodic distortion of the Bi-sublattice.

\subsection{\label{sec:results_b}Transport Characterization}

\begin{figure}[h!]
 \includegraphics[width = 85mm]{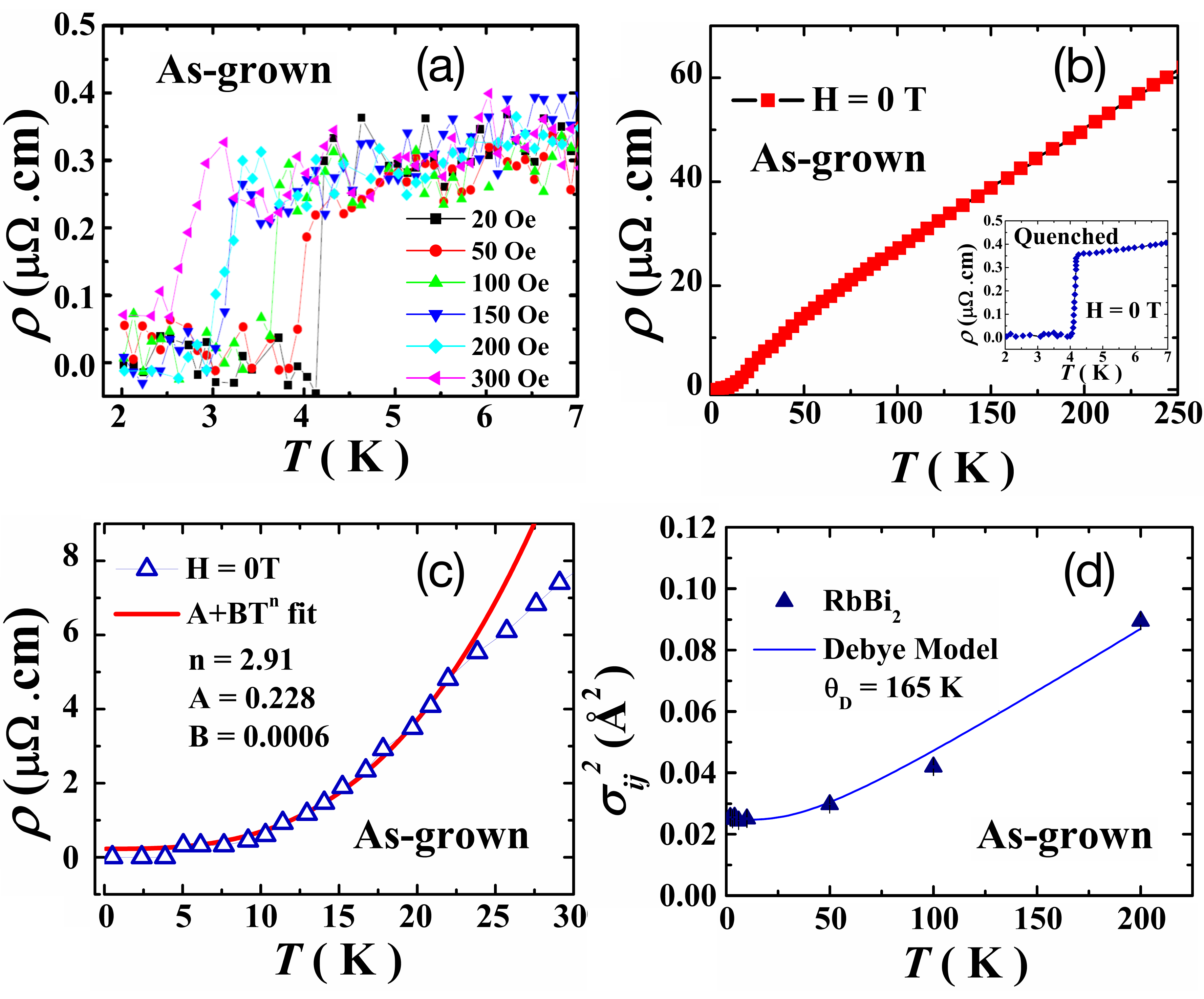} 
\caption[width 
 = \textwidth]{(a) The electrical resistivity showing superconducting transition in the presence of applied field for  RbBi\textsubscript{2} below the critical field. (b) The resistivity as a function of temperature for AG sample of RbBi\textsubscript{2} in the absence of magnetic field. The resistivity of the quenched sample is shown in the inset. (c) The low temperature resistivity fitted with power law relation with temperature $\rho = A + BT^n$ with $n=2.91$. The solid line shows the fitting. (d) The temperature dependence of squared FWHM values of the first peak of the local structural data of RbBi\textsubscript{2}, extracted from a Gaussian fitting. The fit of correlated Debye model is represented by the solid line. All the measurements were done on the AG sample.}
\label{fig:Characterization_electronic_a}
\end{figure}

The results from the transport measurements under magnetic field and temperature on the AG RbBi\textsubscript{2} samples are summarized in Fig.~\ref{fig:Characterization_electronic_a}. Shown in Fig.~\ref{fig:Characterization_electronic_a}(a) is the electrical resistivity as a function of temperature under an applied magnetic field for the AG RbBi\textsubscript{2} confirming the superconducting nature of the ground state. The superconducting transition is sharp with the transition width $\Delta T_c$ $\sim$ 0.41 K at a $T\textsubscript{c}$ of 4.12 K. The superconducting transition shifts to lower temperatures in the presence of larger applied magnetic fields below the critical field.  Shown in Fig.~\ref{fig:Characterization_electronic_a}(b) is the resistivity as a function of temperature in the absence of applied field. At zero field,  RbBi\textsubscript{2} exhibits very good metallic conductivity even at room temperature. The resistivity follows a linear temperature dependence until about 25 K, below which the temperature dependence becomes nearly cubic as it approaches the superconducting transition. The linear-in-temperature dependence of the resistivity upon warming is dominated by electron-phonon scattering which yields $\rho$ $\propto$ $T$. At low temperatures, the resistivity follows a power law temperature dependence. The power law fitting $\rho = A + BT^n$ on resistivity data below 25 K is shown in  Fig.~\ref{fig:Characterization_electronic_a}(c) with $n=2.91$. Similar behavior has been seen in metals such as Pb in which electron-phonon scattering dominates above $T\textsubscript{c}$ ($\sim$ 7.2 K)~\cite{eiling1981pressure}. 
The same transport measurements were carried out for the Q RbBi\textsubscript{2} with the same results. 

\begin{figure}[h!]
 \includegraphics[width = 85mm]{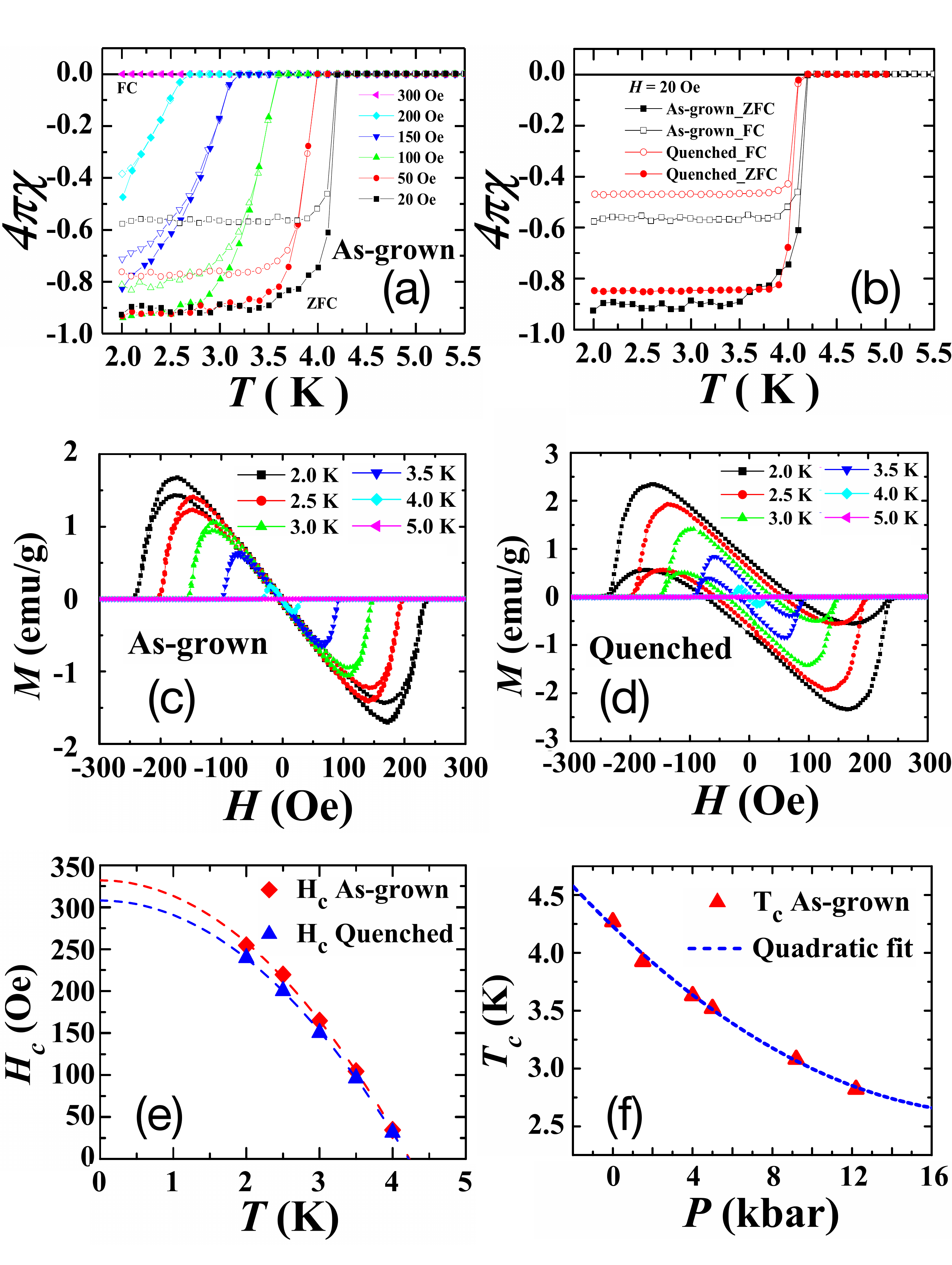} 
\caption[width 
 = \textwidth]{The magnetic characterization of AG ((a), (b), (c), (e)) and Q ((b), (d), (e)) RbBi\textsubscript{2}. Shown in (a) is the bulk susceptibility, $\rchi$,  for the AG sample as a function of field. Shown in (b) is the comparison of bulk susceptibility for the AG and Q samples.  Shown in (c) and (d) are magnetization curves, M(H), after the demagnetization correction for the AG and Q samples, respectively. Very little hysteresis is observed in the AG M(H) unlike in the Q sample. Shown in (e) is the critical magnetic fields of AG and Q samples of RbBi\textsubscript{2} obtained from the magnetization isotherms. The dashed lines represent a fit to the data. Shown in (f) is the behavior of superconducting transition temperature of AG RbBi\textsubscript{2} in the presence of applied pressure. The dashed line shows the quadratic fit on the pressure dependence of $T\textsubscript{c}$.}
\label{fig:Characterization_magnetic}
\end{figure}

\subsection{\label{sec:results_c}Magnetization Characterization}
The results from the magnetic characterization are summarized in Fig.~\ref{fig:Characterization_magnetic}  for the AG and Q samples of RbBi\textsubscript{2}. The low temperature DC magnetic susceptibility measurements exhibit diamagnetic signals with sharp transitions appearing at $T_c$ = 4.15 K ($\Delta T_c$ = 0.30 K) seen in both zero field cooled (ZFC) and field cooled (FC) curves (Fig.~\ref{fig:Characterization_magnetic}(a)). The superconducting transition temperature at 20 Oe is in agreement with that obtained from the transport measurement, confirming the superconducting ground state of RbBi\textsubscript{2} below $T_c$. In the AG sample, 95 \% shielding fraction is observed and a Meissner fraction of 60 \% is determined upon FC for fields that are well below $H_{c}(0)$ (Fig.~\ref{fig:Characterization_magnetic}(b)). In the Q sample, the shielding fraction is 85 \% and the Meissner fraction is 45 \%. This reflects the presence of more pinning centers that trap magnetic flux lines and are not fully expelled when the sample becomes superconducting. With increasing the magnetic field, the superconducting transition gradually shifts to lower temperatures and the transition width becomes wider. The magnetization ($M$) as a function of applied magnetic field, $M(H)$, was measured at various temperatures and was used to estimate the value of the critical field $H_c(T)$. While the superconducting behavior and transition temperature remain the same in both the AG and Q samples of RbBi\textsubscript{2}, AG RbBi\textsubscript{2} shows a full superconducting volume fraction with a sharp hysteresis in M(H) (Fig.~\ref{fig:Characterization_magnetic}(a),(c)) as opposed to Q RbBi\textsubscript{2} which shows a partial volume fraction and a wide hysteresis(Fig.~\ref{fig:Characterization_magnetic}(b),(d)).

The magnetic characterization results identify RbBi\textsubscript{2} to be a type-I superconductor with critical field $H_{c}(0)$ = 332 Oe as shown in Fig. 4(e).  The critical field for the Q sample was determined to be $H_{c}(0)$ = 308 Oe. For comparison, superconductivity in KBi\textsubscript{2} single crystals was recently studied and reported to be a Type-I Bardeen–Cooper–Schrieffer (BCS) superconductor in the dirty limit based on the behavior of magnetization isotherms and the low value of the Ginzburg-Landau (GL) parameter~\cite{sun2016type}. Lastly, shown in Fig.~\ref{fig:Characterization_magnetic}(f) is the pressure dependence of the superconducting transition temperature. $T_{c}$ falls off quadratically with hydrostatic pressure. By 12 kBar, $T_{c}$ drops by nearly 50 \%.  Typically, in metals the percentage change in $T_{c}$ in a similar pressure range is small (examples include In and Pb)~\cite{seiden1969pressure}. Exceptions include Zn and Cd where a nearly 50$\%$ drop in $T_{c}$ has been reported~\cite{seiden1969pressure}. Pressure affects the lattice parameter, Debye temperature and phonon frequencies and $T_c$ $\sim$ $\theta_D*e^{-1/\lambda}$ where $\lambda$ is the electron-phonon coupling constant~\cite{seiden1969pressure}. The large response to pressure is consistent with the low Debye temperature ($\sim$165 K) which renders the lattice very soft. The Debye temperature was determined from the width of the first PDF peak using a correlated Debye model, shown in Fig.~\ref{fig:Characterization_electronic_a}(d). The electron-phonon coupling constant, $\lambda$, was estimated to be 0.734 using McMillan's theory. A non-linear pressure dependence of $T_{c}$ may be due to a change of the Fermi surface topology, from a closed to an open surface. Such a transition causes a change of the density of states as well as the superconducting gap ~\cite{makarov1965anomalies,lifshitz1960high}. Although $T_{c}$ does not change between the AG and Q samples, the superconducting volume fraction and M(H) curves show differences between the two. Slow cooling rates (AG) provide for a full shielding fraction with perfect diamagnetism which is the ideal limit. Fast cooling rates (Q) provide for a reduced shielding fraction. The kagome lattice distortions described above may be linked to the different diamagnetic susceptibility behaviors observed between the two samples.

\section{\label{sec:conclusion}Conclusion}
RbBi\textsubscript{2} is a type-I superconductor with peculiar characteristics. Two RbBi\textsubscript{2} samples were studied, AG and Q, to investigate the effects of structural distortions, electrical resistivity and pressure dependence on $T_{c}$. The critical temperature $T_c$ is 4.15 K. The magnetic and transport properties characterized under field indicate that RbBi\textsubscript{2} is an extremely good metal above $T_c$. The diamagnetic susceptibility for the Q sample does not reach -1 while it does for the AG that may be due to the quenched disorder that breaks the symmetry of the Bi hexagon, creating flux pinning centers and possibly due to differences in grain boundaries. The Q sample has a much broader magnetic hysteresis loop which can be interpreted in terms of the presence of an increase in pinning centers. The neutron diffraction experiments were carried out on Q and AG samples at temperatures above and below the superconducting transition. In the AG sample, the Bi hexagon is undistorted, while in the Q sample, small Bi displacements lead to a distorted Bi hexagon ring. In this, the Bi-Bi bond lengths are not equivalent and distortions of the Bi-rings are spread out in real space. This may be linked to flux pinning and thus one of the reasons for reduced diamagnetism observed in the Q sample.

\section{Acknowledgements}
This work has been supported by the Department of Energy (DOE), Grant number DE-FG02-01ER45927. A portion of this research used resources at the Spallation Neutron Source, a DOE Office of Science User Facility operated by Oak Ridge National Laboratory. Work at Los Alamos National Laboratory was performed under the auspices of the U.S. Department of Energy, Office of Basic Energy Sciences, Division of Materials Science and Engineering.

\bibliography{RbBi2}
\end{document}


\author{Sharon S. Philip}
 \altaffiliation{University of Virginia}
\author{Junjie Yang}
\altaffiliation{NJIT}
\author{Joe Thompson}
\altaffiliation{CMMS, Los Alamos National Laboratory}

\author{Priscila F. S. Rosa}
\altaffiliation{CMMS, Los Alamos National Laboratory}

\author{Katharine L. Page}
\altaffiliation{ORNL}

\author{Despina Louca}
 \email{louca@virginia.edu}
\altaffiliation{University of Virginia}
\title{Supplemental materials to the manuscript *****}
\maketitle

\section{Pair distribution function analysis at low $r$}

The G(r)  pair  correlation  functions  corresponding to the local atomic structures for the AG and Q samples can be modelled using the cubic structure. Fig.~\ref{fig:TemperaturecomparePDF} shows the neutron PDF up to 50{\AA} for the as-grown sample of RbBi\textsubscript{2} at temperature 2K and 200K fitted using the Cubic-Lave phase with $Fd\overline{3}m$ symmetry. The overall agreement with data and the fit in this $r$ range is very good and confirms that there is little disorder in the local structure up to 50{\AA} at both 2K and 200K temperatures. The small differences in local structure arise beyond 50{\AA} and becomes more evident at higher temperature as shown in the main text. 

\begin{figure}[h!]
  \includegraphics[width=8cm]{Figures/Supp_Fig/2and200AG_2to40.pdf}
  \caption{G(r) data and structure fit using $Fd\overline{3}m$ symmetry for as-grown sample of RbBi\textsubscript{2} at 2K and 200K.}
  \label{fig:TemperaturecomparePDF}
\end{figure}

\section{Band structure calculation}

\begin{figure}[h!]
 \includegraphics[width = 80mm,scale=0.5]{Figures/Supp_Fig/Topologicalquantumchem_RbBi2_bandstructure.pdf} 
 \caption{The band structure and density of states(DOS) calculated with spin-orbit coupling(SOC) for RbBi\textsubscript{2}.}
 \label{fig:BandstructureRbBi2}
\end{figure}


\section{Transport behavior and Kohler's scaling}

\begin{figure}[h!]
 \includegraphics[width = 80mm,scale=0.5]{Figures/Supp_Fig/Kohlers_fit_0T1T.pdf} 
  \caption{The temperature dependence of $\rho$ at applied field H = 9T fitted with Kohler scaling $\rho(T,9) =\rho_0(T) + \alpha*B^m/\rho_0^{m-1}$ with $\rho_0$ at (a) H = 0T and (b) H = 1T. The fit parameters were calculated to be $\alpha = 0.951 , m = 1.12$ and $\alpha = 0.333 , m = 5.68$ for (a) and (b) respectively. This suggests that the resistivity behavior at H = 9T for RbBi\textsubscript{2} does not strictly follow Kohler scaling. The temperature dependence in the Kohler's relation 
comes from $\rho_0$. Inset in (a) and (b) shows temperature dependence of $\rho_0$ and the power law functional behavior of best fit. RbBi\textsubscript{2} undergoes a superconducting transition at 4.25K and $\rho$ drops to zero. The fitted curve has been extrapolated to zero below this temperature to account for this offset in (a). At H =1T, RbBi\textsubscript{2} is metallic down to 2K, the critical field being .}
 \label{fig:Kohlersfit}
\end{figure}